%% file: pdferr.tex
\documentclass[12pt]{JHEP3}

\def\HW{\textsc{\small HERWIG}}
\def\pythia{\textsc{\small PYTHIA}}
\def\hpp{\mbox{\textsf{Herwig++}}}
\def\SHERPA{\textsf{SHERPA}}
\def\qmax{{\tilde q_{\rm max}}}
\def\qmin{{\tilde q_{0}}}
\def\q{{\tilde q}}
\def\I{{\cal I}}
\def\P{{\cal P}}

\def\as{{\alpha_S}}

\usepackage{epsfig, bm, amsmath}

\title{Uncertainties of Sudakov form factors} 
\author{Stefan Gieseke\\
  Institut f\"ur Theoretische Physik\\
  Universit\"at Karlsruhe, 76128 Karlsruhe, Germany
  \email{gieseke@particle.uni-karlsruhe.de}
}
\abstract{We study the uncertainties of Sudakov form factors as the
  basis for parton shower evolution in Monte Carlo event generators.
  We discuss the particular cases of systematic uncertainties of parton
  distribution functions and scale uncertainties.}


\preprint{KA-TP-10-2004\\
  hep-ph/0412342}

\begin{document}
\section{Introduction}

The simulation of final states in the collision of highly energetic
particles at current and future colliders is largely based upon parton
shower evolution models.  The theoretical foundation of these models
is the interpretation of Sudakov form factors as no--emission
probabilities \cite{MarchesiniWebber, sjostrand}.  For hadronic
collisions, the space--like evolution of initial state partons from
the hard scale $Q_h$ of the partonic subprocess and relatively small
$x$ is 'guided' by the ratio of parton distribution functions (pdf's)
\cite{book}.  In this paper we would like to address the effect of
uncertainties in the Sudakov form factors as a result of pdf
uncertainties \cite{MRST2001, CTEQ6}.  We shall compare this effect to
the uncertainty due to the choice of scale in the strong coupling
constant $\as$.  We choose to use the Sudakov form factors that are
going to be used for the initial state evolution in future versions of
\hpp{} \cite{hpp10}.  The conclusion should qualitatively remain the
same for the Sudakov form factors that results from different
kinematical variables in the programs \HW{} \cite{Herwig65} and
\pythia{}\cite{Pythia6} or the new program \SHERPA{} \cite{SHERPA}.
They may, however, differ in details as the different kinematics will
lead to emphasis on different regions of evolution phase space.

Furthermore we limit our discussion to the case of spacelike
evolution.  The estimated size of uncertainties due to effects like
scale variation in $\as(Q)$ should be similar in the case of timelike
evolution.  As the pdf's only appear in spacelike evolution we can
study all effects of our interest here. Another limitation is the
discussion of the uncertainties due to terms in the Sudakov form
factor that are formally subleading in the sense of the
next--to--leading (soft and/or collinear) logarithmic approximation.
Formally we may add or subtract terms in the splitting functions that
give rise to only non--logarithmic terms after integration over $\q$
(see below).  Adjusting these terms can be crucial for matching of
matrix elements and parton showers in the CKKW scheme \cite{CKKW,
  CKKWhadron} as discussed in \cite{MrennaPeter, sherpaWjets}.  The
effect, however, appears to be mostly of technical importance close to
the matching scale inherent to this kind of matching scheme.

\section{Sudakov form factor for space--like branchings}

The Sudakov form factor for spacelike backward evolution of a parton
$a$ from the hard scale $\tilde q_{\rm max}$ down to some scale
$\tilde q$ can be written as \cite{sjostrand, book}
\begin{equation}
  \label{eq:sudakov}
  S_a(\q, \qmax; x, \qmin) 
  = \exp\left[-\sum_b\I_{ba}(\q, \qmax; x, \qmin)\right]\;.
\end{equation}
The sum on the right hand side (rhs) is over all possible splittings
into partons of type $b$ and
\begin{equation}
  \label{eq:Idef}
  \I_{ba}(\q, \qmax; x, \qmin) = 
  \int_{\q^2}^{\qmax^2} \frac{d\q^2}{\q^2} \int_{z_0}^{z_1} dz
  \frac{\as(z, \q^2)}{2\pi} 
  \frac{x'f_b(x', \q^2)}{xf_a(x, \q^2)}
  P_{ba}(z, \q^2)\;.
\end{equation}

\FIGURE{
  \input{kin.pstex_t}
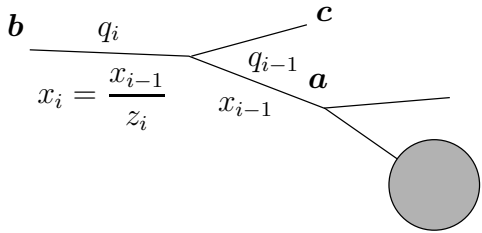
  \caption{Kinematics for space--like branching $b\to ac$.}
  \label{fig:spacebranch}
} 

We assume a resolution scale $\qmin$ below which the evolution
terminates and further splittings would remain unresolvable.  The
limits of the $z$--integration, $z_0$ and $z_1$ depend implicitly on
$\q$ and $\qmin$, hence the $\qmin$ dependence{}\footnote{Some authors
  prefer to stress this circumstance by writing $S_{ba}$ as a ratio of
  two exponentials with explicit $\qmin$ dependence in the
  $\q$--integration.  The two notations are, of course, equivalent.}
of $\I_{ba}$ and $S_a$.  $x$ is the light cone momentum fraction of
the parent parton with respect to the originating hadron.  This value
is initially selected by the hard subprocess.  $x' = x/z$ is the light
cone momentum fraction of the new parent after the first space like
branching and so forth.  $P_{ba}(z, \q^2)$ is the unregularized
collinear splitting function, which, in the case of the evolution of
massive partons, may also depend on the branching scale $\q^2$.  Note,
that the splitting function is regularized as we use explicit cutoffs
in the phase space for soft gluon emission.  $f_a(x, \q^2)$ is the
parton distribution function (pdf) of a parton of type $a$ inside a
proton\footnote{The conclusions will remain unchanged when considering
  antiprotons.  The case of pions or resolved photons in the initial
  state is beyond the scope of this study.}.

The integral in Eq.~\eqref{eq:Idef} is written in a fairly symbolic
way and holds for different models.  We shall rewrite it for the
specific case of the kinematic variables that are used in \hpp{}
\cite{NewVariables}.  There, the evolution variables $z, \q$ are
interpreted in a Sudakov basis.  Consider the branching in
Fig.~\ref{fig:spacebranch}.  Each momentum is written as
\begin{equation}
  \label{eq:sudbas}
  q_i = \alpha_i p + \beta_i n + q_{\perp i}
\end{equation}
where $p$ and $n$ are suitable forward and backward light like vectors,
typically chosen along the axis of the original hadrons.  Then we
define our space--like branching kinematics via 
\begin{equation}
  \label{eq:zqdef}
  \alpha_i = \frac{\alpha_{i-1}}{z}, 
  \qquad 
  \bm{q}_{\perp i} = \frac{\bm{q}_{\perp i-1} - \bm{p}_{\perp i}}{z_i}\;.
\end{equation}
Here, the relative transverse momentum $\bm{p}_{\perp i}$ is given in
terms of the evolution variables $\q$ and $z$ as 
\begin{equation}
  \label{eq:pperpdef}
  \bm{p}_{\perp i}^2 = (1-z_i)^2 \q_i^2 - z_i Q_g^2\;.
\end{equation}
$Q_g$ is the cutoff parameter of the parton shower.  Choosing the
argument of $\as(Q)$ as $Q=(1-z_i)\q_i$ we may now rewrite the
integral \eqref{eq:Idef} as
\begin{equation}
  \label{eq:Idefnew}
  \I_{ba}(\q, \qmax; x, Q_g) = 
  \int_{\q^2}^{\qmax^2} \frac{d\q^2}{\q^2} \int_{0}^{1} dz
  \frac{\as[(1-z)\q]}{2\pi} 
  \frac{x'f_b(x', \q^2)}{xf_a(x, \q^2)}
  P_{ba}(z, \q^2)\Theta(\mathrm{P.S.})\;.
\end{equation}
\FIGURE{
  \includegraphics{psplots.0}
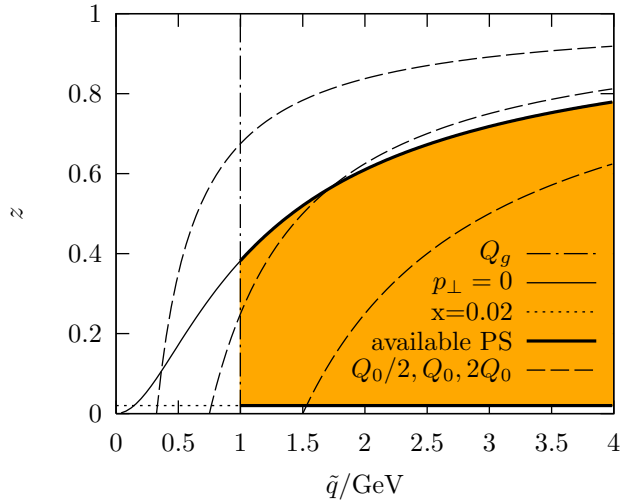
  \caption{Available phase space for space--like branching.}
  \label{fig:phase}
} 
The implicit introduction of the parton shower cutoff $Q_g$ in
\eqref{eq:pperpdef} is reflected in the step function
$\Theta(\mathrm{P.S.})$.  Hence we simply extend the integration
region in $z$ from $0$ to $1$ and apply this cut on the available
parton shower phase space (see also Fig.~\ref{fig:phase}).  However,
the resolution scale $\qmin$ of the parton shower has disappeared and
was implicitly replaced by $Q_g$.  Now every given initial condition
$\qmax, x$ may result in different values of the resolution scale in
terms of a smallest available scale.  The shape of the available phase
space is, of course, still determined by $Q_g$ in some intuitive way
such that small cutoffs lead to a larger available phase space and a
lower resolution scale.  We set $Q_g\approx 1\,$GeV if not stated
otherwise\footnote{More precisely we choose the \hpp{} parameter
  $\delta=2.3\,$GeV (cf.\ \cite{hpp10}).  This leads to $Q_g$ close to
  but slightly smaller than $1\,$GeV.}.

In practice, the available phase space is determined by some more
factors.  As we require a real transverse momentum we find the limits
of the $z$ integration for a given $\q$.  As $x'$ should never exceed
1, $x$ itself is the lower limit of $z$.  Therefore, 
\begin{equation}
  \label{eq:zlimits}
  x < z < 1+\frac{Q_g}{2\q} - \sqrt{\left(1+\frac{Q_g}{2\q}\right)^2 -1}\;.
\end{equation}

Depending on the choice of how to evaluate $\as(Q)$ when $Q$ becomes
very small we have an additional constraint on the phase space.  Let
$Q_0$ be the scale where non--perturbative effects become important.
Typically $Q_0$ is somewhat larger than but of the order of
$\Lambda_{\rm QCD}$.  In order to model non--perturbative effects, we
may simply freeze the value of $\as$, $\as(Q<Q_0)=\as(Q_0)$. We set
$Q_0 = 0.75\,$GeV and freeze $\as$ by default.  This gives us a
behaviour of $\as$ very similar to the one in the MRST2001
parametrization \cite{MRST2001} as we find it in \cite{LHAPDF}.  If we
put $\as(Q<Q_0)=0$ we have an additional limitation of phase space.
In Fig.~\ref{fig:phase} we show as an illustrative example the
available phase space for $q\to qg$ spacelike branching with
$\qmax=4\,$GeV and $x=0.02$ with its limiting regions.  Lines of
constant $\as$ for $Q_0/2, Q_0, 2Q_0$ are also shown.

We apply an extra cut at $\q < Q_g$ in the rare cases where there is
still phase space available.  As the pdf's become unstable for $Q$
values close to the non--perturbative domain we choose to freeze the
pdf's at $Q_{\rm freeze} = 2.5\,$GeV.  The choice of this parameter is
based on experience with \HW{}\footnote{In \HW{} the pdf's are
  regularized by the parameter \texttt{QSPAC} with the default value
  2.5\,GeV.}.

In the following we consider the Sudakov form factor for specific
branching types $b\to ac$ only. Formally we rewrite \eqref{eq:sudakov}
as
\begin{equation}
  \label{eq:sudprod}
  S_a(\q, \qmax; x, Q_g) = \prod_{b} S_{ba}(\q, \qmax; x, Q_g)
\end{equation}
where $S_{ba} = \exp -\I_{ba}$.  Then we only consider specific
$S_{ba}$.  For the parton shower evolution the Sudakov form factor is
interpreted as the probability that the parton $a$ remains unresolved
upon evolution from $\qmax$ down to $\q$ when the minimal resolution
is $\qmin$.  With the probability $1-S_{ba}$ that there is any
branching between $\qmax$ and $\q$ we find the branching probability
density
\begin{equation}
  \label{eq:prob}
  \P_{ba}(\q, \qmax; x, Q_g) = \I_{ba}'(\q; x, Q_g) 
  S_{ba}(\q, \qmax; x, Q_g)\;,
\end{equation}
where $\I_{ba}'(\q; x, Q_g)$ is the integrand of \eqref{eq:Idefnew}
with respect to the $\q$ integration.  The scales of the next
branching $\q$ are distributed according to $\P_{ba}(\q, \qmax; x,
Q_g)$.  This is the quantity we would like to consider in order to
estimate uncertainties of different types as it is directly
implemented in parton shower Monte Carlo programs.  We should note
that $\P_{ba}(\q, \qmax; x, Q_g)$ is normalized to  
\begin{equation}
  \label{eq:norm}
  \int_{\qmax}^{\q_0} \P_{ba}(\q, \qmax; x, Q_g) \,d\q = 
  1-S_{ba}(\qmin, \qmax; x, Q_g)\;.  
\end{equation}
This is typically different from $1$ as we have a chance
$S_{ba}(\qmin, \qmax; x, Q_g)$ to find no branching at all.  Their sum
is obviously one.

\section{Numerical study}
\label{sec:numerical}

For numerical purposes we map the integration variables $\q$ and $z$
onto new variables $t, y$ as 
\begin{equation}
  \label{eq:mapping}
  t = \ln \q^2\;, \qquad
  y = \ln \frac{z}{1-z} \;,
\end{equation}
in order to absorb the leading poles into the integration variables, 
\begin{equation}
  \label{eq:differentials}
  dt = \frac{d\q^2}{\q^2}\;, \qquad
  dy = \frac{dz}{z(1-z)}\;.
\end{equation}
We carry out the integrations in $\I_{ba}'(\q; x, Q_g)$ and
$\I_{ba}(\q, \qmax; x, Q_g)$ numerically with adaptive Gaussian or
Monte Carlo algorithms \cite{NRC, GSL, CUBA}.  Different packages have been
used for consistency checks.  Particular care has been taken to obtain
numerical results with a typical relative error of at least one order
of magnitude below the resulting pdf errors as the former ones would
mask the latter ones otherwise.

In the following we address the uncertainties of Sudakov form factors
that arise from various sources.  As we do not aim at a study of
Sudakov form factors beyond leading logarithmic accuracy, the sources
of uncertainties are:
\begin{enumerate}
\item the ratio of pdf's, which may even be pdf's of different flavour in
  the case of $q\to gq$ and $g \to q\bar q$ branching,   
\item parametrization of the running of $\as$ may play an
  important r\^{o}le,
\item modelling of $\as$ in the non--perturbative region below
  our choice of $Q_0$.   
\end{enumerate}
Recently, the uncertainties of parton distribution functions have been
considered for various observables and the question has been raised
how much of an influence they would have on Monte Carlo showering
algorithms \cite{Joey}.

In order to assess the relevance of uncertainties arising from the
parton distribution functions we basically follow the procedure
described in \cite{MRST2001} and \cite{CTEQ6}. We obtain the integral
once for the central pdf fit and once for all the error members.  The
variations from each pair of error pdf's are added in quadrature and
thus we obtain an error estimate $\delta\I_{ba}$ for the integral from
which we can derive the uncertainty in the probability distribution
\ref{eq:prob}.  All calculations have been performed with the pdf
error distributions from MRST and CTEQ \cite{MRST2001, CTEQ6} as they
are provided in \cite{LHAPDF}.

\FIGURE{ \includegraphics[scale=0.9]{psplots.6}
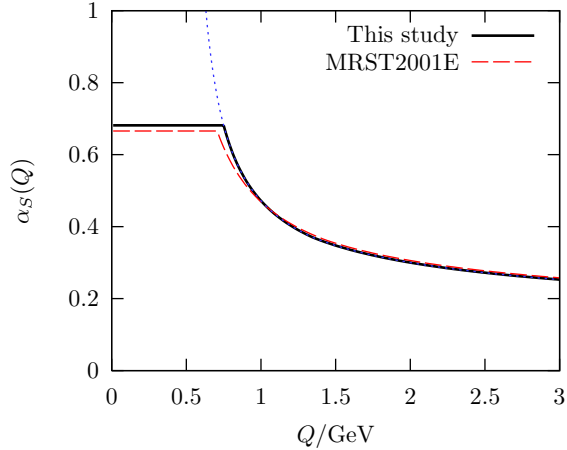
  \caption{Different parametrizations of $\as(Q)$.}
  \label{fig:alphas}
} The parametrization of the running of $\as(Q)$ may give important
differences in the result.  Particularly as different orders of the
running result in large deviations at smaller values of $Q$.  In
Fig.~\ref{fig:alphas} we show our default parametrization with and
without freezing as well as the MRST parametrization of $\as(Q)$.  By
default we use the two--loop $\as$ that is implemented in \hpp. We
choose to freeze $\as$ at $Q_0=0.75\,$GeV as discussed above.  This
parametrization is numerically very close to that of MRST.  The
parametrization of the $\as$ in the CTEQ distribution is very similar
but does not model $\as$ at small scales $Q$ in some particular way.
We therefore do not check this parametrization explicitly.  For MRST
we compared results from a computation with the given pdf set where by
default we use our own parametrization of $\as$ with the same
computation where the $\as$ is taken from the distribution of the pdf
set itself.

As the results seemed to be fairly sensitive on the choice of
$\as$--\-pa\-ra\-me\-tri\-zation we have also checked the effect of
setting $\as$ to zero below $Q_0$.  This may have drastic consequences
as one can see with the help of Fig.~\ref{fig:phase}.  There, we
plotted lines of constant $Q_0$ once for the central value and once
the values resulting from rescaling by factors of 2.  We see that we
may cut out regions in phase space that actually give very large
contributions to the integral $\I_{ba}$.  For estimates of this scale
uncertainty we compute the Sudakov form factors and the branching
probability densities with variation of the scale by a factor two.

\section{Results}
\label{sec:results}
We consider the Sudakov form factors and branching probability
densities in the following.  We choose different initial conditions
$\qmax, x$ for the evolution and focus on a particular type of
branching at a time.  The plots contain three panels and show
$S_{ba}$ on top, $\P_{ba}$ in the middle and some error information
on the bottom panel.

\subsection{Different initial conditions}
\label{sec:ini}

We first consider different initial conditions and branching types.
In this case we always use MRST partons.  We plot Sudakov from factor
and branching probability density for a number of different initial
conditions.  The error box in the lowest panel always compares the
error from the numerical integration with the total error that also
contains the pdf error itself.  This is to control that the numerical
error is negligible compared to the pdf error.  This is always the
case as the numerical error is typically two orders of magnitude
smaller than the pdf error.

\FIGURE[p]{ 
  \parbox{7.4cm}{\includegraphics[scale=0.8]{sudakov3.0}}
  \parbox{7.4cm}{\includegraphics[scale=0.8]{sudakov3.3}}\\[0.5cm]
  \parbox{7.4cm}{\includegraphics[scale=0.8]{sudakov3.8}}
  \parbox{7.4cm}{\includegraphics[scale=0.8]{sudakov3.11}}
  \caption{  \label{fig:plots1}
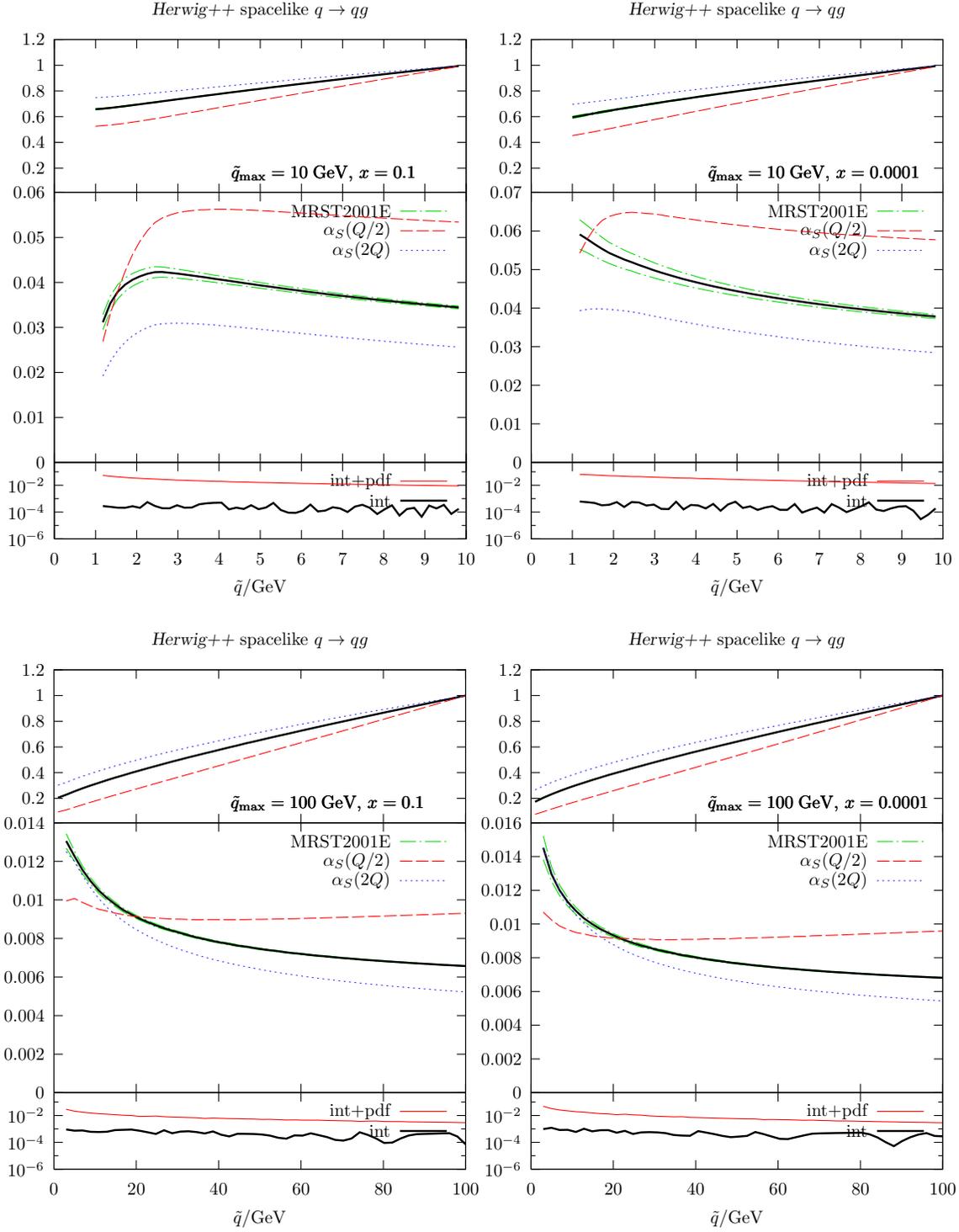
    pdf and $\as$ uncertainties in spacelike $q\to qg$ branchings
    for different initial conditions $\qmax$ and $x$. Each plot shows
    $S_{ba}(\q)$ (top), $\P_{ba}(\q)$ (middle) and the relative error
    $\delta\P_{ba}/\P_{ba}$ from numerical integration and pdf
    uncertainties (bottom panel). The central value is shown as solid
    line, central $\pm$ pdf error is dot--dashed and $\as$
    uncertainties from rescaling the argument by a factor $1/2$ and
    $2$ are shown as dashed/dotted line, respectively.}  }

In Fig.~\ref{fig:plots1} we show the uncertainties in the case of
$q\to qg$ spacelike branching.  We pick $d$ quarks in particular.
Clearly, the estimated pdf error is by far smaller than the one from
varying the $\as$ scale by a factor of two.  Generally, the pdf
uncertainty is largest at small initial scales $\qmax = 10\,$GeV and
somewhat larger at small $x$.  This goes along with the fact that the
quark densities are known better at large $x$ than at small $x$.  We
do not show any plots at intermediate values of $x$ and $\q$ as they
don not exhibit any new features.  The plots shown in
Fig.~\ref{fig:plots1} are those with a large visible effect.

\FIGURE[p]{ 
  \parbox{7.4cm}{\includegraphics[scale=0.8]{sudakov3.36}}
  \parbox{7.4cm}{\includegraphics[scale=0.8]{sudakov3.38}}\\[0.5cm]
  \parbox{7.4cm}{\includegraphics[scale=0.8]{sudakov3.39}}
  \parbox{7.4cm}{\includegraphics[scale=0.8]{sudakov3.43}}
  \caption{\label{fig:plots2}
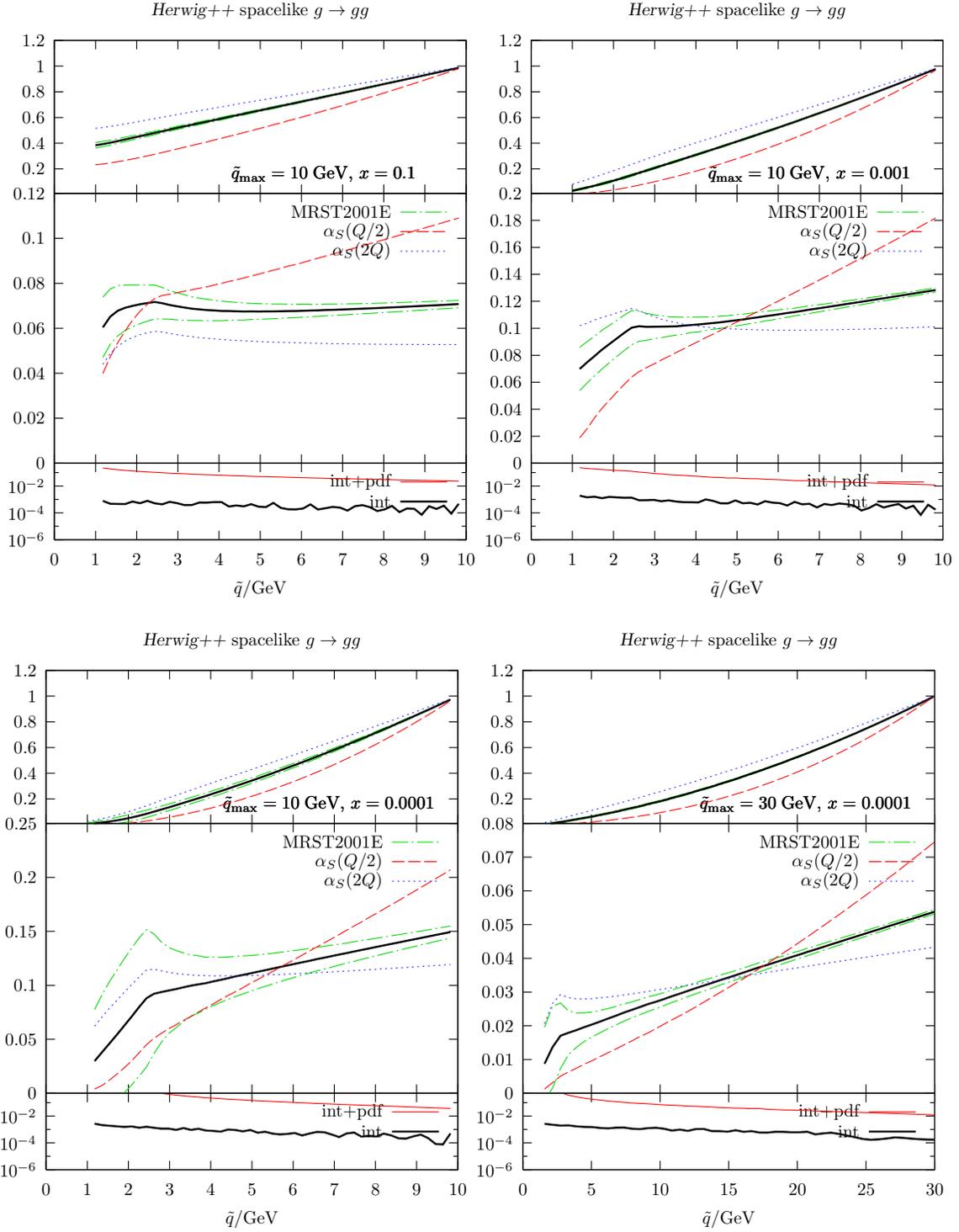
    pdf and $\as$ uncertainties in spacelike $g\to gg$ branchings at
    different $x$ and $\qmax$ values. See caption of
    Fig.~\ref{fig:plots1} for labelling.}}

For $g\to gg$ branchings we get a slightly different picture in
Fig.~\ref{fig:plots2}. The pdf uncertainties are much larger in this
case as the gluon density is not as much constrained as the valence
quark density.  This is particularly true at large $x$.  Going to
small $x=10^{-4}$ at $\qmax=10\,$GeV the pdf uncertainty is very large
(the largest we encounter in this study) and sometimes even larger than
the uncertainty from $\as$ scale variation.  This is still the case if
one goes to a larger initial $\qmax = 30\,$GeV but the effect is still
largest at small target values of $\q$. For even larger $\qmax$ (not
shown) we observe the same trend.

\FIGURE[p]{ 
  \parbox{7.4cm}{\includegraphics[scale=0.8]{sudakov3.12}}
  \parbox{7.4cm}{\includegraphics[scale=0.8]{sudakov3.15}}\\[0.5cm]
  \parbox{7.4cm}{\includegraphics[scale=0.8]{sudakov3.24}}
  \parbox{7.4cm}{\includegraphics[scale=0.8]{sudakov3.27}}
  \caption{  \label{fig:plots3}
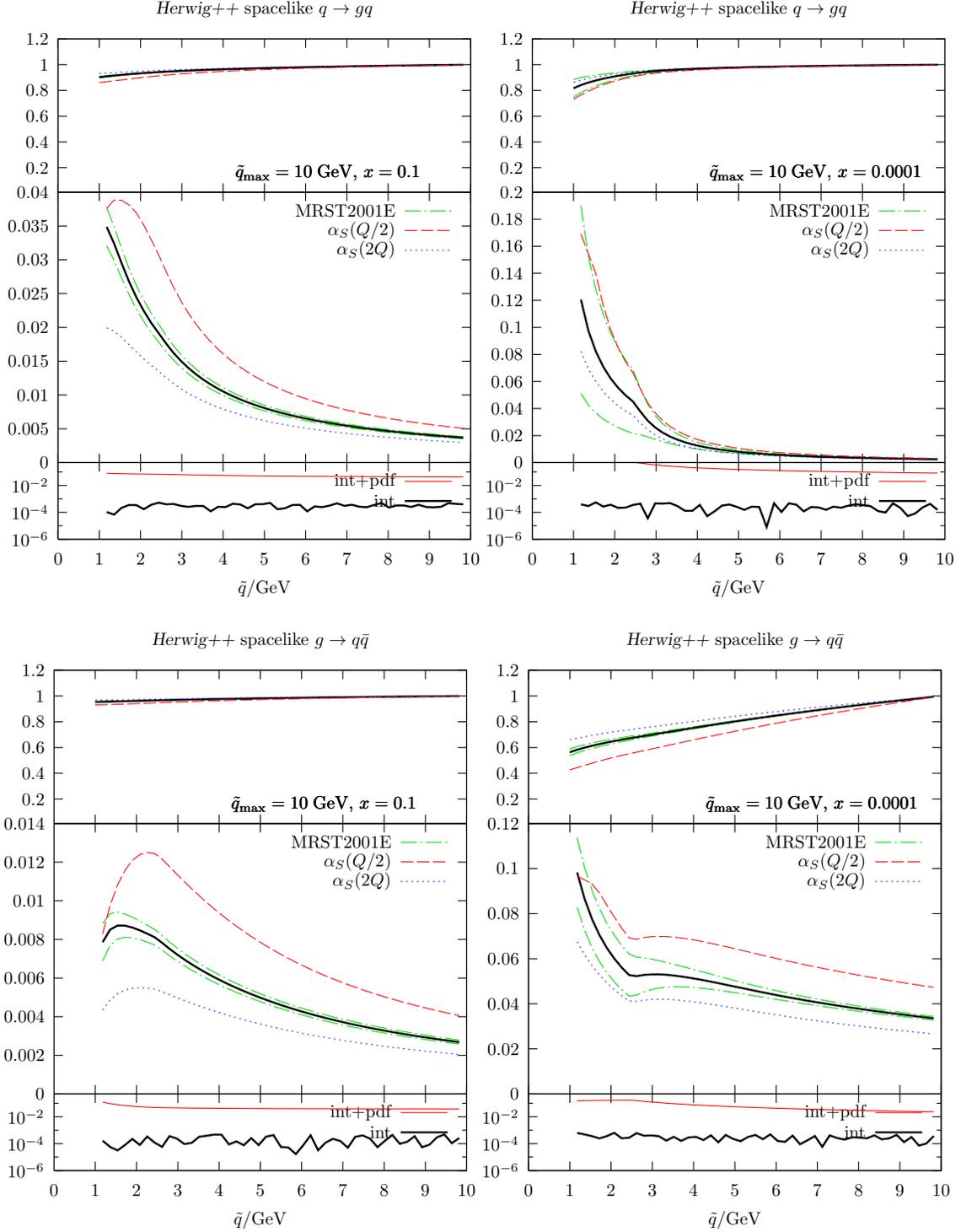
    pdf and $\as$ uncertainties of the branchings $q\to gq$
    and $g\to q\bar q$ at different $x$--values. See caption of
    Fig.~\ref{fig:plots1} for labelling.}
}

We also consider the more unlikely branchings $g\to gq$ and $g\to
q\bar q$ in Fig.~\ref{fig:plots3}.  The Sudakov form factor for these
branchings is typically close to one.  The pdf uncertainties can be
sizable as well, particularly when we involve a gluon at small values
of $x$ where we also had the largest uncertainty in the case of $g\to
gg$ splitting.  Still, we have selected situations with particularly
large pdf uncertainties.  Generally, we have the larger error from
$\as$ scale variation. 

\FIGURE[p]{ 
  \parbox{7.4cm}{\includegraphics[scale=0.8]{sud_anti.0}}
  \parbox{7.4cm}{\includegraphics[scale=0.8]{sud_anti.2}}\\[0.5cm]
  \parbox{7.4cm}{\includegraphics[scale=0.8]{sud_anti.12}}
  \parbox{7.4cm}{\includegraphics[scale=0.8]{sud_anti.24}}
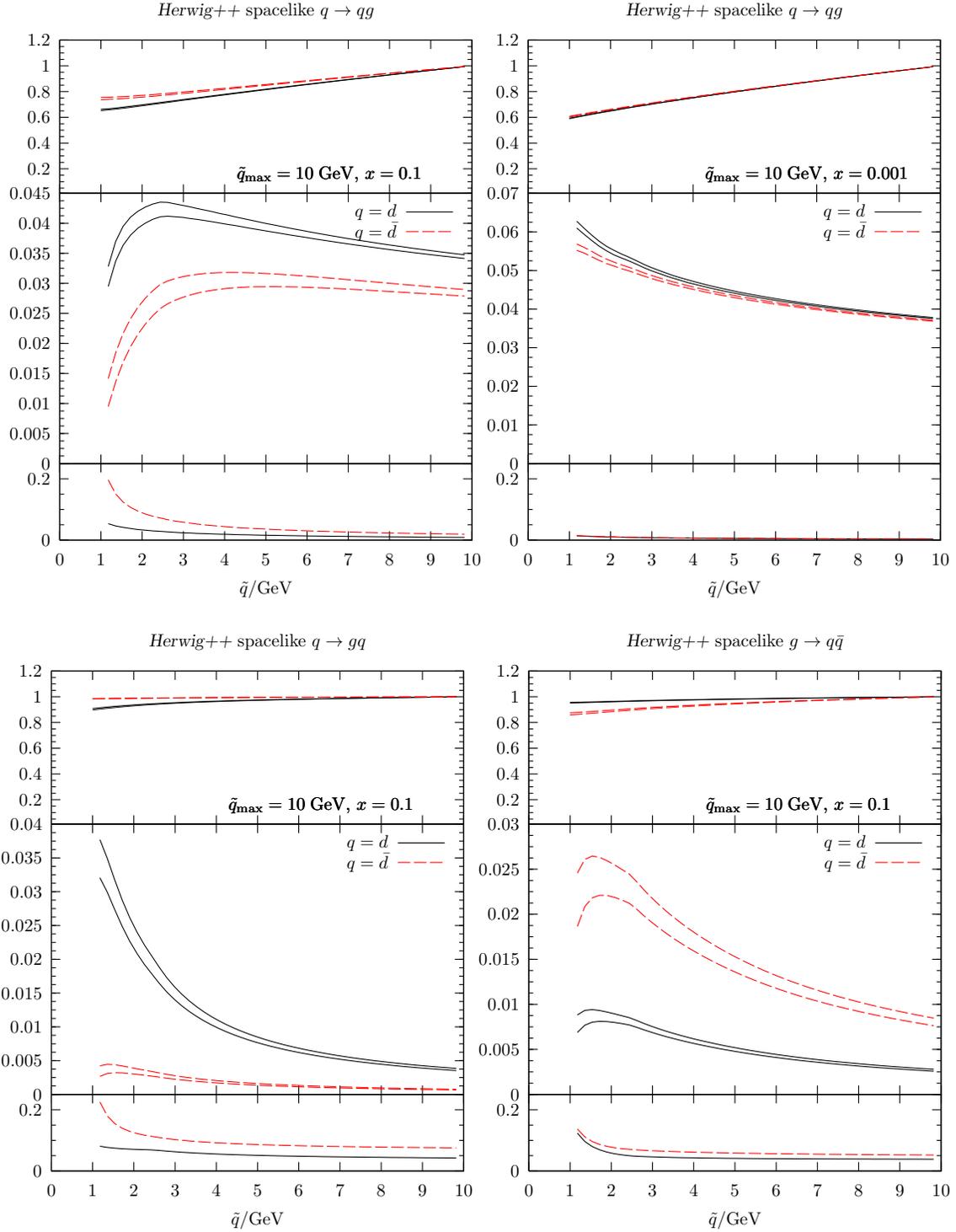
  \caption{\label{fig:plots4} Comparison of uncertainties in the
    evolution of valance ($d$) quarks and sea ($\bar d$) quarks. The
    lowest panels show the total relative error of the branching
    probability densities.  The solid lines are partly identical to
    the pdf error bands in Figs.~\ref{fig:plots1}
    and~\ref{fig:plots3}. See caption of Fig.~\ref{fig:plots1} for
    labelling.}}

In Fig.~\ref{fig:plots4} we replace the initial valence quark $a=d$
with a sea quark $a=\bar d$.  Now we can directly observe and compare
the sea--like behaviour of the $\bar d$ as opposed to the valence
$d$--quark.  Especially at larger $x=0.1$ the evolution clearly
prefers valence quarks (note, that the branching probability density
is normalized to the total probability of such a branching to occur).
We can directly compare to the picture at small $x=10^{-3}$ where the
two distributions begin to coincide as the valence structure loses its
dominance. The opposite happens of course for evolution \emph{away}
from the $d/\bar d$ into a gluon.  The evolution of the $d$ is clearly
suppressed.   As far as the errors due to pdf's are concerned we
observe that the uncertainty of the sea quark density at large $x$
becomes visible as it does for the gluon.    This is highly emphasised
at the low end of the evolution. 

\subsection{MRST vs CTEQ}
\label{sec:mrstcteq}

\FIGURE[p]{ 
  \parbox{7.4cm}{\includegraphics[scale=0.8]{sudakov4.15}}
  \parbox{7.4cm}{\includegraphics[scale=0.8]{sudakov4.24}}\\[0.5cm]
  \parbox{7.4cm}{\includegraphics[scale=0.8]{sudakov4.36}}
  \parbox{7.4cm}{\includegraphics[scale=0.8]{sudakov4.39}}
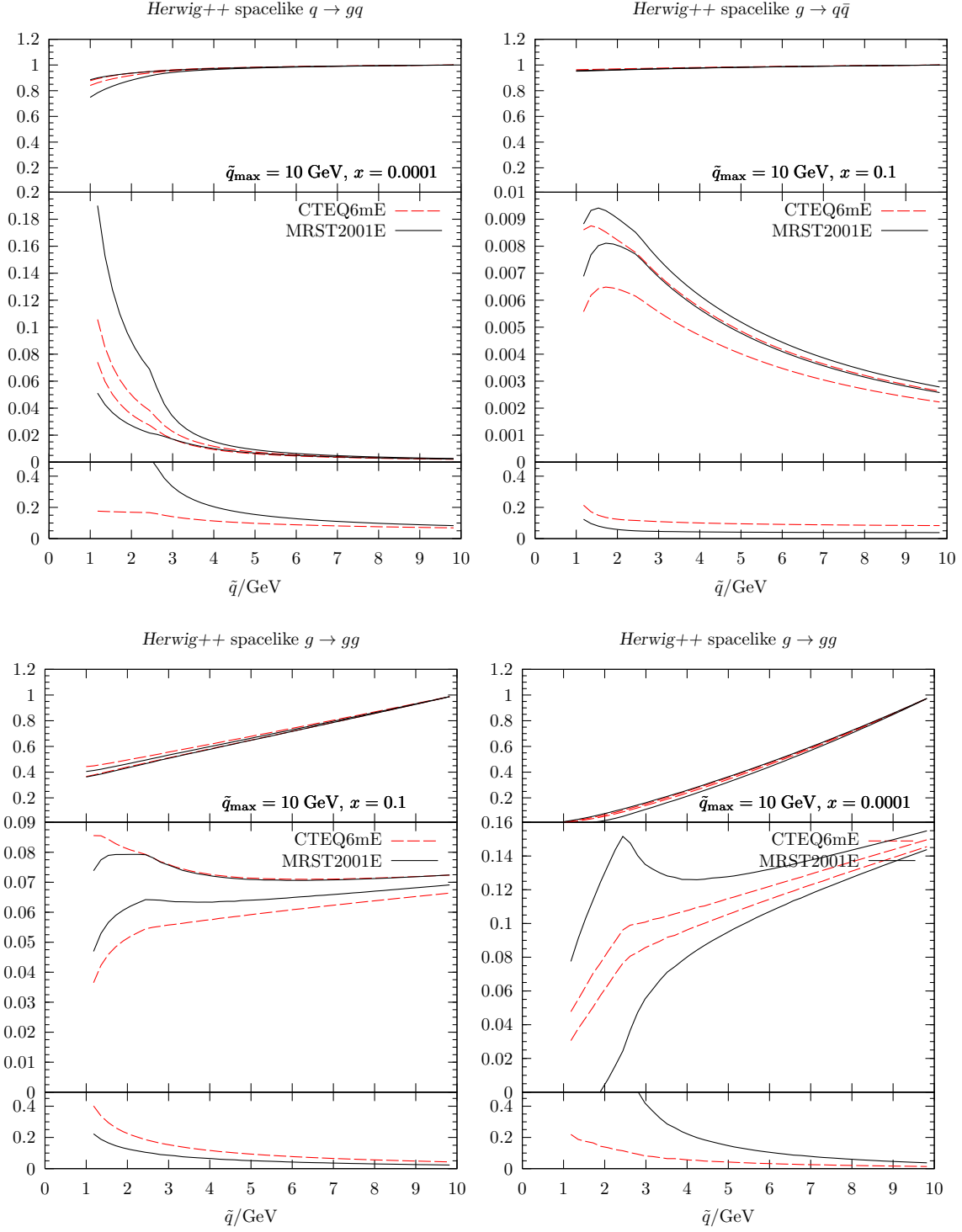
  \caption{Comparison of pdf uncertainties from MRST2001E (solid) and CTEQ6mE
    (dashed) parametrizations where differences are most significant.
    Upper and middle panel are similar to Fig.~\ref{fig:plots1}.  The
    bottom panel shows the relative pdf error of the branching
    probability density. See caption of Fig.~\ref{fig:plots1} for
    labelling.\label{fig:plots5}} }

As an obvious cross check we repeated all computations also with the
CTEQ6m error set.  In Fig.~\ref{fig:plots5} we show some of the
previous plots again, without $\as$ uncertainties, comparing the error
bands of the two parton distribution packages.  We find that the
distributions are compatible but the error estimates from CTEQ are
smaller, particularly for gluons at small $x$.  Again, we have picked
cases that show clear differences.  Generally, both packages tend to
give similar (and compatible) results.

\subsection{Parametrization of \boldmath $\as$}
\label{sec:aspar}

\FIGURE[p]{ 
  \parbox{7.4cm}{\includegraphics[scale=0.8]{sud_mrst.0}}
  \parbox{7.4cm}{\includegraphics[scale=0.8]{sud_mrst.4}}
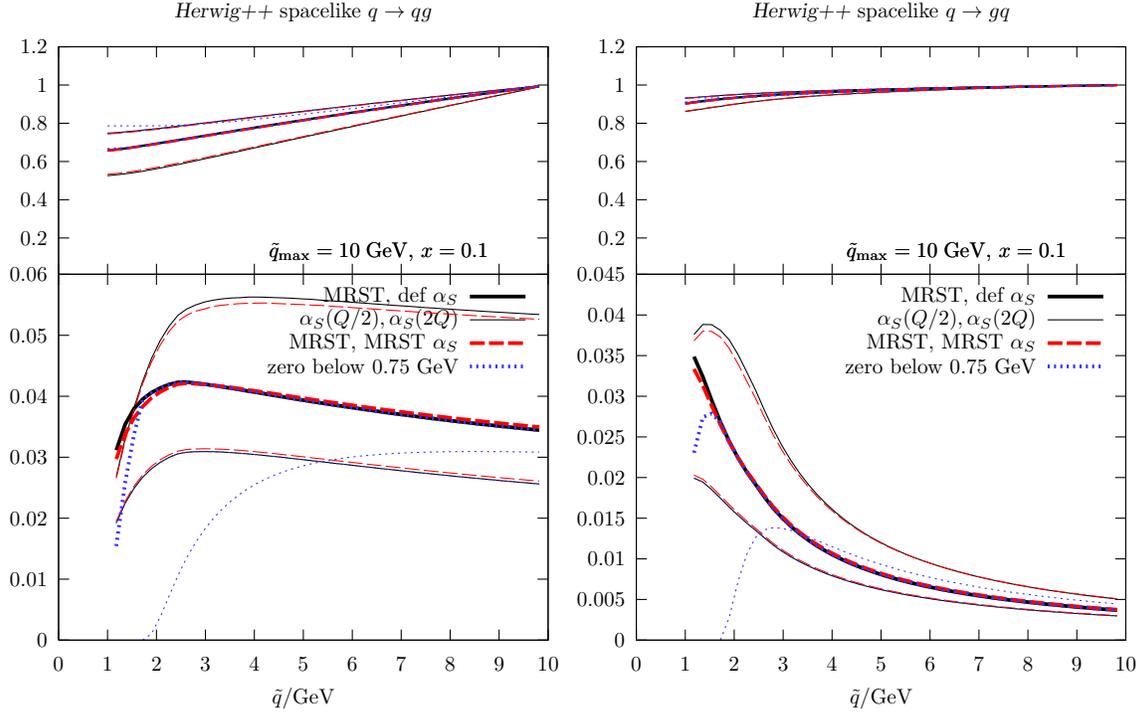
  \caption{Effect of different $\as$ parametrizations in spacelike
    $q\to qg$ and $q\to gq$ branchings. The solid line in both plots
    denotes the previous results, obtained with MRST partons and our
    default parametrization of $\as$, including the effect of varying
    the scale by a factor of 2. The dashed and dotted lines are
    results obtained with different parametrizations of $\as$.}
  \label{fig:plots6}
}

As already pointed out in the discussion of the available phase space,
the effect of different parametrizations of $\as(Q)$, especially at
low $Q$ can have important effects.  In Fig.~\ref{fig:plots6} we show
the effect of using the $\as$ parametrization that is provided via the
distribution of MRST ifself with our parametrization.  We have chosen
a very similar regularisation for non--perturbative values of $\as$ as
MRST have.  Therefore the results are very similar, nevertheless they
are sinsitive to the small differences in the parametrization of
$\as$.  However, the effect of using a parametrization that sets $\as$
to zero below e.g.\ $Q_0=0.75\,$GeV leads to a very different result
as effectively a part of phase space is modified in which $\I_{ba}$
picks up very large contributions.  This is of particular importance
when we use $Q/2$ as this cuts out a lot of phase space.  In the other
two cases the results closely follow those of our default
parametrization until they hit the phase space boundary.

\subsection{Larger cutoff}
\label{sec:cutoff}

\FIGURE[p]{ 
  \parbox{7.4cm}{\includegraphics[scale=0.8]{sud_top.1}}
  \parbox{7.4cm}{\includegraphics[scale=0.8]{sud_top.3}}
  \parbox{7.4cm}{\includegraphics[scale=0.8]{sud_top.10}}
  \parbox{7.4cm}{\includegraphics[scale=0.8]{sud_top.12}}
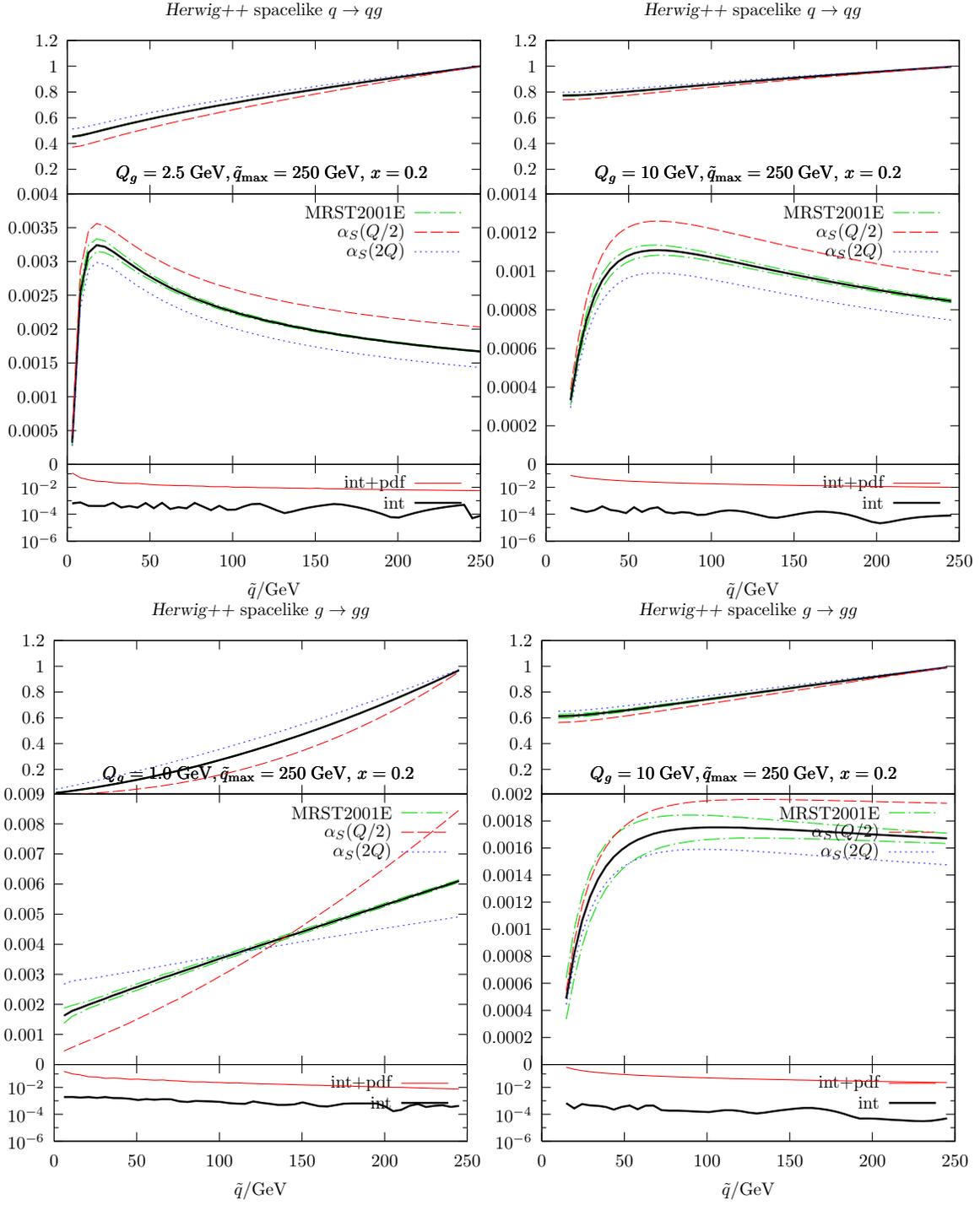
  \caption{Increasing the cutoff $Q_g$ in the $q\to qg$ and $g\to gg$
    Sudakov form factors for at large $\qmax$ and $x$.  See caption of
    Fig.~\ref{fig:plots1} for labelling.}
  \label{fig:plots7}
}

As a final subject we consider the spacelike evolution of light quarks
or gluons at large $x=0.2$ and fairly high initial scale
$\qmax=250\,$GeV.  This situation can give us some information about
the initial state radiation in top production at hadron colliders.
$t\bar t$ pairs are predominantly produced from light quark scattering
at large $x$.  The relevant initial scale $\qmax$ is given by a
typical invariant mass in the $t$--channel as the colour connection is
between one initial light quark and one $t$--quark
\cite{NewVariables}.  This gives us a typical initial scale $\qmax =
250\,$GeV.

In Fig.~\ref{fig:plots7} we have plotted the relevant Sudakov form
factors and branching probability densities for such a situation.  We
have increased the cutoff scale to higher values $Q_g=2.5\,$GeV and/or
10\,GeV.  This allows us to directly estimate the effect of the
radiation of an extra gluon at a fairly large (visible) scale.  We
find that the pdf uncertainties remain fairly low in the Sudakov form
factors themselves.  The size of the branching probability
distribution is affected to some extent but the shape is hardly
modified.  Again, the uncertainty from the variation of $\as$ is much
larger.

\section{Conclusion}
\label{sec:conclusion}

We have studied the effect of various sources of uncertainties on the
parton shower evolution.  We therefore particularly considered the
branching probability density as this is the quantity that drives the
evolution.  We generally find that the effect of pdf uncertainties is
small compared to uncertainties in the QCD description of the
evolution itself that we estimate via scale variation in the strong
coupling.  There are, however, particular regions in phase space where
the pdf uncertainties are very large.  This is the case for the
evolution from very small values of $x$ and also at large $x$ when sea
quarks or gluons are involved.  Considering the initial conditions in
$\q$ we find that the pdf uncertainties are generally small when the
initial conditions are in the large $\q$ regime.  In the course of the
parton shower evolution, however, all showers have to evolve through
the small $\q$ in some way. Besides, we emphasise that the evolution
strongly depends on the modelling of $\as$ in the transition region
between perturbative and non--perturbative QCD.  This uncertainty is,
however, usually taken as a sensitivity that is exploited in tuning
Monte Carlo showers to experimental data.

In summary, we find that one should be aware of pdf uncertainties in
the evolution when one is confined to a phase space of fairly large
$x$ and relatively small initial $\q$.

\acknowledgments 

I would like to thank J.\ Huston for fruitful conversations and for
raising this topic at the HERA/LHC workshop.  I also thank Peter
Richardson, Mike Seymour and Bryan Webber for valuable comments.

\end{document}

%% file: kin.pstex_t
\begin{picture}(0,0)%
\includegraphics{kin.pstex}%
\end{picture}%
\setlength{\unitlength}{4144sp}%
\begingroup\makeatletter\ifx\SetFigFont\undefined%
\gdef\SetFigFont#1#2#3#4#5{%
  \reset@font\fontsize{#1}{#2pt}%
  \fontfamily{#3}\fontseries{#4}\fontshape{#5}%
  \selectfont}%
\fi\endgroup%
\begin{picture}(2834,1395)(-134,-694)
\put(406,479){\makebox(0,0)[lb]{\smash{{\SetFigFont{12}{14.4}{\familydefault}{\mddefault}{\updefault}$q_i$}}}}
\put(1666,164){\makebox(0,0)[lb]{\smash{{\SetFigFont{12}{14.4}{\rmdefault}{\mddefault}{\updefault}$\bm a$}}}}
\put(1126, 29){\makebox(0,0)[lb]{\smash{{\SetFigFont{12}{14.4}{\rmdefault}{\mddefault}{\updefault}$x_{i-1}$}}}}
\put(1306,299){\makebox(0,0)[lb]{\smash{{\SetFigFont{12}{14.4}{\rmdefault}{\mddefault}{\updefault}$q_{i-1}$}}}}
\put(-134,479){\makebox(0,0)[lb]{\smash{{\SetFigFont{12}{14.4}{\rmdefault}{\mddefault}{\updefault}$\bm b$}}}}
\put(1711,569){\makebox(0,0)[lb]{\smash{{\SetFigFont{12}{14.4}{\rmdefault}{\mddefault}{\updefault}$\bm c$}}}}
\put( 46, 74){\makebox(0,0)[lb]{\smash{{\SetFigFont{12}{14.4}{\rmdefault}{\mddefault}{\updefault}$\displaystyle x_i=\frac{x_{i-1}}{z_i}$}}}}
\end{picture}%